\documentclass{svproc}
\usepackage{url}

\usepackage{amsmath}
\usepackage{amssymb}
\usepackage[misc]{ifsym}
\usepackage{graphicx}
\usepackage{graphics}
\usepackage{xcolor}

\begin{document}
\mainmatter              % start of a contribution
\title{A space--dependent Boltzmann--BGK model for gas mixtures and its hydrodynamic limits}
\titlerunning{Boltzmann--BGK kinetic model}  % abbreviated title (for running head)
%                                     also used for the TOC unless
%                                     \toctitle is used
%
\author{Marzia Bisi\inst{1} \and Maria Groppi\inst{1} \and Giorgio Martal\`o\inst{1}}
\authorrunning{M. Bisi et al.} % abbreviated author list (for running head)
%
%%%%% list of authors for the TOC (use if author list has to be modified)
%\tocauthor{Ivar Ekeland, Roger Temam, Jeffrey Dean, David Grove,
%Craig Chambers, Kim B. Bruce, and Elisa Bertino}
%
\institute{$^1$Department of Mathematical, Physical and Computer Sciences,\\ University of Parma, Italy, \vspace*{0.1 cm}\\
%$^2$Department of Mathematics F. Casorati,\\ University of Pavia, Italy \vspace*{0.1 cm}\\
{\it Corresponding author:} \email{marzia.bisi@unipr.it}\\
}

\maketitle              % typeset the title of the contribution

\begin{abstract}
We present a hybrid Boltzmann-BGK model for inert mixtures, where each kind of binary interaction may be described by a classical Boltzmann integral or by a suitable relaxation-type operator. We allow also the possibility of changing the option Boltzmann/BGK operator according to the space position. We prove that this model guarantees conservations of species masses, global momentum and energy, as well as the entropy dissipation, leading to the expected Maxwellian equilibria with all species sharing the same mean velocity and the same temperature. We investigate then such mixed kinetic equations in three different hydrodynamic limits: the classical collision dominated regime, a situation with dominant intra--species collisions, and a mixture with heavy and light particles leading to a kinetic--fluid description.

\end{abstract}
\section{Introduction}

The kinetic approach to mixtures of rarefied gases is based on a set of evolution equations for the distribution function of each constituent. The basic properties of the multi--species Boltzmann system may be found in \cite{Cercignani2,Chapman-Cowling}, but such integro--differential equations are still under investigation, both from the analytical and the numerical points of view \cite{Bondesan-etal,Gamba-Colic}.
Many consistent relaxation models of BGK--type have been also built up for gas mixtures (even for polyatomic or reactive constituents) \cite{AAP,BBGSP,Bisi-Caceres,GSPoF,HaackHauckMurillo}, and their simple structure makes them more suitable for numerical applications and to derive closed fluid--dynamic systems from kinetic equations, at different levels of accuracy. In the paper \cite{LucchinKRM} a hybrid Boltzmann-BGK model has been proposed for mixtures of monatomic gases, that combines the detailed description of collisions given by the Boltzmann integral operators with the simplicity and the numerical manageability of BGK-type relaxation operators.
That kinetic model has the same structure of the full Boltzmann equations, with the collision term of each constituent given by a sum of bi-species operators, that may be chosen either of Boltzmann or of BGK type. Each binary BGK operator has the form prescribed in \cite{BBGSP}, and preserves exchanges of momentum and energy in each binary interaction of the corresponding Boltzmann operator. The mixed Boltzmann-BGK model turns out to reproduce correct conservation laws and Maxwellian equilibria, and fulfills the Boltzmann H-theorem.

In this paper, we propose and investigate a space--dependent Boltzmann--BGK model, where the option of choosing a Boltzmann or a BGK description may change according to the space position. This model allows thus for the possibility of making use of the detailed Boltzmann operators in regions showing rapid changes in the trend of distribution functions (as the formation of shock wave profiles) and of the simpler BGK approximation in regions where the configuration is not far from a steady state. The space--dependent coefficients $\chi_{ij}({\bf x}) \in \{ 0,1 \}$ determine the choice of a Boltzmann operator (for $\chi_{ij}({\bf x})=1$) or a BGK one (for $\chi_{ij}({\bf x})=0$) for collisions between species $i,j$ in the position ${\bf x} \in \mathbb{R}^3$; for any choice of these coefficients, we prove that the hybrid kinetic model is consistent and preserves the entropy dissipation and the correct Maxwellian equilibria, with all species having the same mean velocity and the same temperature.
The presence of a collision operator for any pair of gaseous components allows for a consistent derivation of evolution equations for the main macroscopic fields in different hydrodynamic regimes, according to the dominant collision process. Besides the regime dominated by the whole collision operator and the case where only intra--species collisions are dominant, discussed and justified also in \cite{LucchinKRM,Boscheri-etal}, we focus the attention here on a mixture of heavy and light particles, where the very disparate masses lead to very frequent collisions among the heavy particles and almost negligible interactions among the light ones, while collisions between heavy and light atoms have an intermediate rate. This scaling leads to macroscopic Euler--type equations for species fields of the heavy constituent, and to a kinetic equation for the light gas; this hydrodynamic limit shows that kinetic--fluid models commonly used in the literature in order to investigate (analytically and numerically) two--phase flows \cite{Cruseillesetal,DegondJinMieussens,GoudonJinetal} may~be rigorously obtained from a purely kinetic system, in a suitable asymptotic regime.

The organization of the paper is the following. In Section 2 we fix notations and we give preliminary well--known results on the Boltzmann description for gas mixtures and on the BGK approximation proposed in \cite{BBGSP}. Then, Section 3 is devoted to our space--dependent mixed Boltzmann--BGK model and to the proof of its consistency properties (preservation of conservation laws, collision equilibria and validity of the H--theorem). In Section 4 hydrodynamic equations at Euler level are derived both in the classical collision dominated regime, in the case with dominant intra--species collisions, and in situations dominated by the heaviest particles. Some concluding remarks and perspectives are given in Section~5.

\section{Preliminaries on multi--species Boltzmann and BGK equations}

A kinetic model for a mixture of $N$ monatomic gases is constituted by a system of evolution equations for the distribution functions $f_i =f_i (\mathbf{x},\mathbf{v},t)$ of each species $i=1,\ldots,N$. Each distribution depends on position $\mathbf{x}\in\mathbb{R}^3$, velocity $\mathbf{v}\in\mathbb{R}^3$ and time $t\in\mathbb{R}_+$, and main species macroscopic fields may be recovered as suitable moments of $f_i$ with respect to the velocity variable ${\bf v}$. Specifically, number density $n_i$, mean velocity ${\bf u}_i$, and temperature $T_i$ (with normalized Boltzmann constant) are provided by
\begin{equation}
n_i = \int_{\mathbb{R}^3} f_i({\bf v})\, d{\bf v}\,, \quad
{\bf u}_i = \frac{1}{n_i} \int_{\mathbb{R}^3} {\bf v}\, f_i({\bf v})\, d{\bf v}\,,  \quad
T_i = \frac{m_i}{3\, n_i} \int_{\mathbb{R}^3} |{\bf v} - {\bf u}_i|^2\, f_i({\bf v})\, d{\bf v}\,,
\end{equation}
where $m_i$ is the particle mass. Global fields as total number density $n$, mass density $\rho$, mixture mean velocity ${\bf u}$, and mixture temperature $T$ are given by
\begin{equation}
\begin{aligned}
n = \sum_{i=1}^N n_i\,,&  \qquad \rho= \sum_{i=1}^N m_i\, n_i\,,  \qquad
{\bf u} = \sum_{i=1}^N m_i\, n_i\, {\bf u}_i\,,  \\
T &= \frac{1}{n} \sum_{i=1}^N n_i\, T_i + \frac{1}{3\,n} \sum_{i=1}^N m_i\, n_i\, |{\bf u}_i - {\bf u}|^2 \,.
\end{aligned}
\end{equation}

The kinetic system may be cast in general form as
\begin{equation}
	\dfrac{\partial f_i}{\partial t}+\mathbf{v}\cdot\nabla_{\mathbf{x}}f_i=\mathcal{Q}_i(\mathbf{f}), \qquad i=1, \dots, N,
\label{kinetic-eq}
\end{equation}
where on the left hand side we have the streaming operator (describing free flow of particles), and on the right hand side a scattering operator taking into account the effects on distribution $f_i$ of particle collisions (depending on all distributions ${\bf f} = (f_1, f_2, \ldots, f_N)$).

In the classical Boltzmann description \cite{Cercignani2}, the collision operator is a sum of binary contributions:
\begin{equation}
\mathcal{Q}_i(\mathbf{f})=\sum_{j=1}^N \mathcal{Q}_{ij}(f_i,f_j)\,,\qquad i =1,\ldots, N\,,
\label{Boltzmann}
\end{equation}
where $\mathcal{Q}_{ij}$ takes into account only collisions between particles of species $(i,j)$, including the case $j=i$. It is an integral--type operator of the form
\begin{equation}
	\mathcal{Q}_{ij}(f_i,f_j)=\int_{\mathbb{R}^3} \int_{\mathbb{S}^2} \sigma_{ij}(|\mathbf{g}|,\hat{\mathbf{g}}\cdot\boldsymbol{\omega})\left[f_i(\mathbf{v}^\prime_{ij})f_j(\mathbf{w}^\prime_{ij})-f_i(\mathbf{v})f_j(\mathbf{w}) \right]d\mathbf{w}d\boldsymbol{\omega}.
	\label{Botlz_term}
\end{equation}
The function $\sigma_{ij}$ denotes the collision kernel depending on the modulus of the pre--collision relative velocity $\mathbf{g}= \mathbf{v}-\mathbf{w}$ and on the angle between directions of pre-- and post--collision relative velocities ($\hat{\mathbf{g}}$ and $\boldsymbol{\omega}\in\mathbb{S}^2$, respectively).
Post--collision velocities $(\mathbf{v}^\prime_{ij},\mathbf{w}^\prime_{ij})$ are related to the pre--collision ones $(\mathbf{v},\mathbf{w})$ as
\begin{equation}
\begin{aligned}
    \mathbf{v}^\prime_{ij}&= \alpha_{ij}\, \mathbf{v}+ \alpha_{ji}\, \mathbf{w} + \alpha_{ji} |\mathbf{v} - \mathbf{w}|\boldsymbol{\omega}\\
    \mathbf{w}^\prime_{ij}&= \alpha_{ij}\, \mathbf{v}+ \alpha_{ji}\,\mathbf{w} - \alpha_{ij}\, |\mathbf{v} - \mathbf{w}|\boldsymbol{\omega}\,,
\end{aligned}
\end{equation}
with the mass ratio defined as $\alpha_{ij}=m_i/(m_i+m_j)$.
Denoting by $\varphi_i({\bf v})$ a general molecular property and by $\langle \cdot, \cdot \rangle$ the usual $L^2$--product, it can be checked, by exchanging pre-- and post--collision velocities, that the weak form of each binary Boltzmann operator (\ref{Botlz_term}) may be cast as
\begin{equation}
\begin{aligned}
     \langle \mathcal{Q}_{ij}(f_i,f_j), \varphi_i({\bf v}) \rangle &= -\, \frac12 \int_{\mathbb{R}^3} \int_{\mathbb{R}^3} \int_{\mathbb{S}^2} \sigma_{ij}(|\mathbf{g}|,\hat{\mathbf{g}}\cdot\boldsymbol{\omega}) \left[ \varphi_i({\bf v}^\prime_{ij}) - \varphi_i({\bf v}) \right] \vspace*{0.2 cm}\\
    &\times\left[f_i(\mathbf{v}^\prime_{ij})f_j(\mathbf{w}^\prime_{ij})-f_i(\mathbf{v})f_j(\mathbf{w})\right]d\mathbf{v} d\mathbf{w}d\boldsymbol{\omega}\,.
\end{aligned}
\label{weak-Bolt}
\end{equation}
Moreover, it is well known that equilibrium states for the set of Boltzmann equations (\ref{kinetic-eq})--(\ref{Boltzmann}) are given by Maxwellian distributions depending on species densities $n_i$, global mean velocity ${\bf u}$ and global temperature $T$:
\begin{equation}
\mathcal{M}_i=\mathcal{M}_i(\mathbf{v};n_i,\mathbf{u},T)=n_i \left(\dfrac{m_i}{2\pi T}\right)^{3/2} \exp \left[-\dfrac{m_i}{2T} \left| \mathbf{v}-\mathbf{u} \right|^2\right]\,.
\label{Maxw-eq}
\end{equation}

Since integral Boltzmann collision operators are quite awkward to deal with, suitable approximations of BGK--type have been proposed in the literature (see for instance \cite{AAP,GSPoF,HaackHauckMurillo,KlingPirnerPuppo}, just to mention few of them). Generally speaking, in this type of models the Boltzmann collision terms are replaced by simpler relaxation--type operators, so that in (\ref{kinetic-eq}) one has
\begin{equation}
\mathcal{Q}_i(\mathbf{f}) = \sum_{j=1}^N \bar{Q}_{ij}(f_i)\,,\qquad \text{with} \qquad  \bar{Q}_{ij}(f_i)=\nu_{ij}(\mathcal{M}_{ij}-f_i)\,;
\label{BGK-model}
\end{equation}
here, $\nu_{ij}$ stands for a suitable relaxation frequency (free parameter to be fit according to the Boltzmann interaction rates), and the attractor $\mathcal{M}_{ij}$ has a Maxwellian shape
\begin{equation}
	\mathcal{M}_{ij}=\mathcal{M}_{ij}(\mathbf{v};n_{ij},\mathbf{u}_{ij},T_{ij})= n_{ij}\left(\dfrac{m_i}{2\pi T_{ij}}\right)^{3/2} \exp\left[-\dfrac{m_i}{2T_{ij}}\left|\mathbf{v}-\mathbf{u}_{ij}\right|^2\right]\,,
	\label{auxil_maxwel}
\end{equation}
and depends on fictitious quantities $n_{ij}$, $\mathbf{u}_{ij}$, $T_{ij}$. There are many ways to determine the disposable auxiliary fields in terms of species densities, velocities and temperatures in order to guarantee the consistency properties, namely that the BGK model preserves correct conservation laws (of species densities, global momentum and energy), Maxwellian collision equilibria, and fulfillment of Boltzmann H--theorem.

Among the many available BGK descriptions for gas mixtures, in the sequel use will be made of the one proposed in \cite{BBGSP}, that preserves the structure of Boltzmann operators (sum of different binary contributions) and is well suited for general intermolecular potentials. It assumes at first that collisions among the same species are described by the usual BGK operator for a single gas proposed in \cite{BGK}, namely
\begin{equation}
	n_{ii}=n_i\,,\qquad
\mathbf{u}_{ii}=\mathbf{u}_i\,,\qquad T_{ii}=T_i\,.
\end{equation}
Then, $n_{ij}=n_i$ even for $j \not=i$ since the mixture is not reactive, and other auxiliary fields $\mathbf{u}_{ij}$ and $T_{ij}$ (for $j \not=i$) are determined imposing that each bi--species BGK operator preserves the corresponding bi--species Boltzmann exchange rates of momentum and energy, namely that
\begin{equation}
\begin{aligned}
    \langle \mathcal{Q}_{ij},1 \rangle &= \langle \bar{Q}_{ij},1 \rangle=0\,, \qquad  \langle\mathcal{Q}_{ij}-\bar{Q}_{ij},\mathbf{v} \rangle=\boldsymbol{0}\,,\\
    &\langle \mathcal{Q}_{ij}-\bar{Q}_{ij},|\mathbf{v}|^2 \rangle =0 \,, \qquad j\neq i\,.
\end{aligned}
	\label{exch_rates}
\end{equation}
Skipping all intermediate computations, that the interested reader may find in \cite{LucchinKRM,BBGSP}, we get
\begin{equation}
	n_{ij}=n_i\,,\qquad j=1,\ldots,N\,, \quad j\neq i\,,
\end{equation}
\begin{equation}
	\mathbf{u}_{ij}=(1-a_{ij})\mathbf{u}_i + a_{ij} \mathbf{u}_j\,,
\end{equation}
\begin{equation}
	T_{ij}=(1-b_{ij})T_i + b_{ij} T_j +\gamma_{ij}|\mathbf{u}_i-\mathbf{u}_j|^2\,,
\label{aux-temp}
\end{equation}
where
$$
\begin{array}{c}
\displaystyle a_{ij}=\dfrac{\lambda_{ij}\, m_j\, n_j}{\nu_{ij}\, (m_i + m_j)}\,, \qquad
\displaystyle b_{ij}=2\,a_{ij}\, \frac{m_i}{m_i+ m_j}\,, \vspace*{0.2 cm}\\
\displaystyle \gamma_{ij}=\dfrac13 m_i a_{ij} \left( \frac{2\, m_j}{m_i + m_j} -a_{ij} \right)\,.
\end{array}
$$
Coefficients $\lambda_{ij}$ are related to the Boltzmann collision kernels as
$$
\lambda_{ij}(|{\bf g}|)\, \hat{\bf g} = \int_{\mathbb{S}^2} (\hat{\bf g} - \boldsymbol{\omega})\, \sigma_{ij}(|{\bf g}|, \hat{\bf g} \cdot \boldsymbol{\omega})\, d\boldsymbol{\omega}\,.
$$
For Maxwell molecules interactions, with $\sigma_{ij}$ independent of $|{\bf g}|$, coefficients $\lambda_{ij}$ are constant; for general intermolecular potentials it is not possible to compute them explicitly, but in \cite{BBGSP} they have been approximated by their value in a suitable average of the relative speed, and they thus depend on macroscopic fields: $\lambda_{ij}(|{\bf g}|) \cong \lambda_{ij}( \bar{g})$, with
\begin{equation}
    \begin{aligned}
        \bar{g} &= \left( \frac{1}{n_i\, n_j} \int_{\mathbb{R}^3} \int_{\mathbb{R}^3} |{\bf v} - {\bf w}|^2\, f_i({\bf v})\, f_j({\bf w})\,  d{\bf v}\, d{\bf w} \right)^{1/2} \\
        &= \left[ 3\, \left( \frac{T_i}{m_i} + \frac{T_j}{m_j} \right) + |{\bf u}_i - {\bf u}_j|^2 \right]^{1/2}.
    \end{aligned}
\label{bar-g}
\end{equation}
Positiveness of coefficients $\gamma_{ij}$ and therefore of auxiliary temperatures $T_{ij}$ is guaranteed for collision frequencies fulfilling the constraints $\nu_{ij} \geq \lambda_{ij}\, n_j /2$ (see Proposition~2 in \cite{BBGSP}).

\section{Space--dependent Boltzmann--BGK model}

We present a hybrid Boltzmann--BGK description that combines the positive features of Boltzmann and BGK formulations, namely the detailed description of collision mechanisms allowed by Boltzmann operators, and the analytical and numerical manageability of BGK terms.
In \cite{LucchinKRM} we have proposed a mixed kinetic model, where each kind of binary interaction $(i,j)$ may be alternatively described by means of a Boltzmann or a BGK operator. In this paper we aim at generalizing such hybrid model, introducing the possibility of varying the option Boltzmann/BGK operator according to the space position. Specifically, we consider the following Boltzmann--BGK model for an inert mixture:
\begin{equation}
\dfrac{\partial f_i}{\partial t}+\mathbf{v}\cdot\nabla_{\mathbf{x}}f_i= \sum_{j=1}^N \Big[ \chi_{ij}({\bf x})\, \mathcal{Q}_{ij}(f_i,f_j) + (1 - \chi_{ij}({\bf x}))\, \bar{Q}_{ij}(f_i) \Big]\,,\quad i =1,\ldots, N\,,
\label{general-mixed}
\end{equation}
where $\mathcal{Q}_{ij}(f_i,f_j)$ is the bi--species Boltzmann operator defined in (\ref{Botlz_term}), and $\bar{Q}_{ij}(f_i)$ is the binary BGK operator provided in (\ref{BGK-model}).

Coefficients $\chi_{ij} \in \{ 0,1 \}$ may depend on the space position; we assume that $\chi_{ij} = \chi_{ji}$, $\forall i,j=1, \dots, N$. The option $\chi_{ij}({\bf x})=1$ means that collisions between particles $(i,j)$ occurring in position ${\bf x}$ are described by a Boltzmann operator, while the option $\chi_{ij}({\bf x})=0$ means that for such collisions we use the BGK operator described in the previous section. The option $\chi_{ij}({\bf x})=1$, $\forall (i,j)$ and $\forall {\bf x} \in \mathbb{R}^3$, reproduces the full Boltzmann model, while the option $\chi_{ij}({\bf x})=0$, $\forall (i,j)$ and $\forall {\bf x} \in \mathbb{R}^3$, allows to recover the BGK model proposed in \cite{BBGSP}. In this mixed model one could also decide to use the detailed Boltzmann description only in some regions of the space domain, for instance where variations of distributions and of trends of macroscopic fields are expected to be more significant for the specific physical problem under consideration. A simple reasonable example could be obtained by dividing the whole space $\mathbb{R}^3$ into two regions $A$, $B$ (such that $A \cup B = \mathbb{R}^3$, with $A$ open set), setting
$$
\chi_{ij}({\bf x}) = \left\{
\begin{array}{ll}
\chi_{ij}^A \qquad & \text{for}\ {\bf x} \in A \vspace*{0.2 cm}\\
\chi_{ij}^B \qquad & \text{for}\ {\bf x} \in B
\end{array}
\right. \qquad \quad \text{with} \quad \chi_{ij}^A,\ \chi_{ij}^B \in \{ 0,1 \}\,,
$$
but the results that will be presented in the sequel hold for whatever decomposition, even with a higher number of regular sub-domains.

We prove now that this hybrid kinetic model is well--posed even in presence of space--dependent coefficients $\chi_{ij}({\bf x})$, namely that conservation laws are fulfilled, and classical Boltzmann H-theorem holds, implying uniqueness of the correct Maxwellian equilibrium configuration.
As concerns conservations of species densities, global momentum and total kinetic energy, we have to show that
\begin{equation}
\sum_{j=1}^N \langle \chi_{ij}({\bf x})\, \mathcal{Q}_{ij}(f_i,f_j) + (1 - \chi_{ij}({\bf x}))\, \bar{Q}_{ij}(f_i), 1 \rangle = 0\,,\quad \forall\, i =1,\ldots, N\,,
\label{cons-1}
\end{equation}
and
\begin{equation}
\sum_{i=1}^N \sum_{j=1}^N \langle \chi_{ij}({\bf x})\, \mathcal{Q}_{ij}(f_i,f_j) + (1 - \chi_{ij}({\bf x}))\, \bar{Q}_{ij}(f_i), \varphi_i({\bf v}) \rangle = 0\,,
\label{cons-2}
\end{equation}
for $\varphi_i({\bf v}) = m_i\, {\bf v}$ and $\varphi_i({\bf v}) = m_i\, |{\bf v}|^2$.
Equations (\ref{cons-1}) trivially hold, since all binary Boltzmann and BGK operators separately preserve mass, i.e.  $\langle \mathcal{Q}_{ij}(f_i,f_j), 1 \rangle$ $ = 0$, $\langle \bar{Q}_{ij}(f_i),1 \rangle=0$, $\forall (i,j)$.
Recalling that the $L^2$ scalar product $\langle \cdot , \cdot \rangle$ has to be performed with respect to the velocity variable ${\bf v}$, therefore coefficients involving $\chi_{ij}({\bf x})$ may be factorized out, equation (\ref{cons-2}) may be rewritten as
\begin{equation}
\begin{array}{l}
\displaystyle \sum_{i=1}^N \Big[ \chi_{ii}({\bf x})\, \langle \mathcal{Q}_{ii}(f_i,f_i), \varphi_i({\bf v}) \rangle + (1 - \chi_{ii}({\bf x}))\, \langle \bar{Q}_{ii}(f_i), \varphi_i({\bf v}) \rangle \Big] \vspace*{0.2 cm} \\
\displaystyle + \frac12 \sum_{i=1}^N \sum_{\substack{j=1\\j\neq i}}^N \Big[ \chi_{ij}({\bf x}) \Big( \langle \mathcal{Q}_{ij}(f_i,f_j), \varphi_i({\bf v}) \rangle + \langle \mathcal{Q}_{ji}(f_j,f_i), \varphi_j({\bf v}) \rangle \Big) \\
\displaystyle +\, (1 - \chi_{ij}({\bf x})) \Big( \langle \bar{Q}_{ij}(f_i), \varphi_i({\bf v}) \rangle + \langle \bar{Q}_{ji}(f_j), \varphi_j({\bf v}) \rangle \Big) \Big] = 0.
\end{array}
\label{cons-2-mod}
\end{equation}
In this formula the second sum, involving contributions due to interactions between different species, has been obtained by exchanging indices $i$ and $j$ and recalling the symmetry property $\chi_{ji}({\bf x}) = \chi_{ij}({\bf x})$. Terms appearing in equation (\ref{cons-2-mod}) vanish separately. Indeed, $\varphi_i({\bf v}) = m_i\, {\bf v}$ and $\varphi_i({\bf v}) = m_i\, |{\bf v}|^2$ are collision invariants for the single--species Boltzmann and BGK operators, therefore the terms in the first line of (\ref{cons-2-mod}) vanish. Concerning the inter--species contributions, the Boltzmann ones vanish because of conservations of momentum and energy in each collision (the analytical result comes from the weak form (\ref{weak-Bolt}) with $\varphi_i({\bf v}) = m_i\, {\bf v}$ and $\varphi_i({\bf v}) = m_i\, |{\bf v}|^2$); for the BGK terms, explicit simple computations provide
$$
\begin{array}{l}
\langle \bar{Q}_{ij}(f_i), m_i\, {\bf v} \rangle + \langle \bar{Q}_{ji}(f_j), m_j\, {\bf v} \rangle =  \vspace*{0.3 cm} \\
\displaystyle = -\lambda_{ij}\dfrac{m_i m_j}{m_i+m_j} n_i n_j(\mathbf{u}_i-\mathbf{u}_j) + \lambda_{ji} \dfrac{m_i m_j}{m_i+m_j} n_i n_j (\mathbf{u}_i-\mathbf{u}_j) = {\bf 0},
\end{array}
$$
$$
\begin{array}{l}
\langle \bar{Q}_{ij}(f_i), m_i\, |{\bf v}|^2 \rangle + \langle \bar{Q}_{ji}(f_j), m_j\, |{\bf v}|^2 \rangle = \vspace*{0.3 cm} \\
\displaystyle = -\, \lambda_{ij}\, \frac{2\, m_i\, m_j}{(m_i + m_j)^2}\, n_i\, n_j \Big[ 3 (T_i - T_j) + (m_i\, {\bf u}_i + m_j\, {\bf u}_j) \cdot ({\bf u}_i - {\bf u}_j) \Big] \vspace*{0.2 cm} \\
\displaystyle + \lambda_{ji}\, \frac{2\, m_i\, m_j}{(m_i + m_j)^2}\, n_i\, n_j \Big[ 3 (T_i - T_j) + (m_i\, {\bf u}_i + m_j\, {\bf u}_j) \cdot ({\bf u}_i - {\bf u}_j) \Big] = 0\,,
\end{array}
$$
where we have taken into account that $\lambda_{ji} = \lambda_{ij}$.

We prove now that our mixed Boltzmann--BGK model fulfills the entropy dissipation (H--theorem). We define the following H--functional
\begin{equation}
\mathcal{H}({\bf f}) = \sum_{i=1}^N \langle f_i, \log f_i -1 \rangle = \sum_{i=1}^N \int_{\mathbb{R}^3} f_i({\bf v})\, \Big( \log f_i({\bf v}) - 1 \Big)\, d{\bf v}\,.
\label{H-func}
\end{equation}
By a direct computation, its time derivative provides
$$
\frac{\partial \mathcal{H}}{\partial t} =  \sum_{i=1}^N \int_{\mathbb{R}^3} \frac{\partial f_i}{\partial t}\, \log f_i({\bf v})\, d{\bf v}\,,
$$
hence from the kinetic system (\ref{general-mixed}) we get
\begin{equation}
\frac{\partial \mathcal{H}}{\partial t} + \nabla_{\bf x} \cdot \Phi[\mathcal{H}] = \mathcal{S}({\bf f})
\end{equation}
where $\Phi[\mathcal{H}]$ is the entropy flux
$$
\Phi[\mathcal{H}] = \sum_{i=1}^N \int_{\mathbb{R}^3} {\bf v}\, f_i({\bf v})\, \Big( \log f_i({\bf v}) - 1 \Big)\, d{\bf v}\,,
$$
and $\mathcal{S}({\bf f})$ is the entropy production
\begin{equation}
\mathcal{S}({\bf f}) = \sum_{i=1}^N \sum_{j=1}^N \int_{\mathbb{R}^3} \Big[ \chi_{ij}({\bf x})\, \mathcal{Q}_{ij}(f_i,f_j) + (1 - \chi_{ij}({\bf x}))\, \bar{Q}_{ij}(f_i) \Big]\,  \log f_i({\bf v})\, d{\bf v}\,.
\end{equation}
In order to prove that $\mathcal{S}({\bf f}) \leq 0$, as expected in each consistent kinetic system, we separate intra--species and inter--species contributions and we collect together the terms due to operators with exchanged indices:
\begin{equation}
\begin{array}{ccl}
\mathcal{S}({\bf f}) & = & \displaystyle \sum_{i=1}^N \chi_{ii}({\bf x}) \langle \mathcal{Q}_{ii}(f_i,f_i), \log f_i \rangle + \sum_{i=1}^N (1 - \chi_{ii}({\bf x}))\, \langle \bar{Q}_{ii}(f_i), \log f_i \rangle \vspace*{0.1 cm} \\
 & + & \displaystyle \sum_{i=1}^N \sum_{\substack{j=1\\j > i}}^N \chi_{ij}({\bf x}) \Big( \langle \mathcal{Q}_{ij}(f_i,f_j), \log f_i \rangle + \langle \mathcal{Q}_{ji}(f_j,f_i), \log f_j \rangle \Big) \vspace*{0.1 cm}\\
 & + & \displaystyle \sum_{i=1}^N \sum_{\substack{j=1\\j > i}}^N (1 - \chi_{ij}({\bf x})) \Big( \langle \bar{Q}_{ij}(f_i), \log f_i \rangle + \langle \bar{Q}_{ji}(f_j), \log f_j \rangle \Big)\,.
\end{array}
\label{Sf}
\end{equation}
It is well known that for the one--species Boltzmann operator  and for the classical one--species BGK approximation one has  $\langle \mathcal{Q}_{ii}(f_i,f_i), \log f_i \rangle \leq 0$ and $\langle \bar{Q}_{ii}(f_i),$ $ \log f_i \rangle \leq 0$ (see \cite{Cercignani2} and \cite{BGK}, respectively). As concerns bi--species Boltzmann contributions, the weak form (\ref{weak-Bolt}) with test functions $\log f_i$ and $\log f_j$ provides
$$
\begin{array}{l}
\displaystyle \langle \mathcal{Q}_{ij}(f_i,f_j), \log f_i \rangle + \langle \mathcal{Q}_{ji}(f_j,f_i), \log f_j \rangle = \vspace*{0.2 cm} \\
\displaystyle = -\, \frac12 \int_{\mathbb{R}^3} \int_{\mathbb{R}^3} \int_{\mathbb{S}^2} \sigma_{ij}(|\mathbf{g}|,\hat{\mathbf{g}}\cdot\boldsymbol{\omega}) \log \left( \frac{f_i(\mathbf{v}^\prime_{ij})f_j(\mathbf{w}^\prime_{ij})}{f_i(\mathbf{v})f_j(\mathbf{w})} \right) \vspace*{0.2 cm}\\
\displaystyle \times\left[\frac{f_i(\mathbf{v}^\prime_{ij})f_j(\mathbf{w}^\prime_{ij})}{f_i(\mathbf{v})f_j(\mathbf{w})} - 1 \right] f_i(\mathbf{v})\, f_j(\mathbf{w})\, d\mathbf{v} d\mathbf{w}d\boldsymbol{\omega} \leq 0\,,
\end{array}
$$
where the sign follows from the properties of the function $(z-1) \log z$. For the bi--species BGK contributions, at first we note (see \cite{BBGSP} for further details) that
$$
\begin{array}{c}
\langle \mathcal{M}_{ij} - f_i, \log f_i \rangle\, \leq\, \langle \mathcal{M}_{ij}, \log \mathcal{M}_{ij} - 1 \rangle - \langle f_i, \log f_i -1 \rangle \vspace*{0.2 cm}\\
 = \langle \mathcal{M}_{ij}, \log \mathcal{M}_{ij} \rangle - \langle f_i, \log f_i \rangle \leq \langle \mathcal{M}_{ij}, \log \mathcal{M}_{ij} \rangle - \langle \mathcal{M}_{ii}, \log \mathcal{M}_{ii} \rangle
\end{array}
$$
where intermediate equality holds since $f_i$ and $\mathcal{M}_{ij}$ share the same density $n_i$, and last inequality takes into account that $\langle f_i, \log f_i \rangle$ takes its minimum at a Maxwellian \cite{Cercignani2}. Consequently, recalling also that
$$
\log \mathcal{M}_{ij} = \log n_i+\dfrac32\log\left(\dfrac{m_i}{2\pi T_{ij}}\right)-\dfrac{m_i}{2T_{ij}}|\mathbf{v}-\mathbf{u}_{ij}|^2,
$$
we get
$$
\begin{array}{l}
\displaystyle \langle \bar{Q}_{ij}(f_i), \log f_i \rangle + \langle \bar{Q}_{ji}(f_j), \log f_j \rangle \vspace*{0.2 cm}\\
\leq \nu_{ij} \langle \mathcal{M}_{ij}, \log \mathcal{M}_{ij} \rangle - \nu_{ij} \langle \mathcal{M}_{ii}, \log \mathcal{M}_{ii} \rangle + \nu_{ji}
 \langle \mathcal{M}_{ji}, \log \mathcal{M}_{ji} \rangle - \nu_{ji} \langle \mathcal{M}_{jj}, \log \mathcal{M}_{jj} \rangle \vspace*{0.2 cm} \\
\displaystyle = -\, \frac32\, \left[ \nu_{ij}\, n_i \log \left( \frac{T_{ij}}{T_i} \right) + \nu_{ji}\, n_j \log \left( \frac{T_{ji}}{T_j} \right) \right] \vspace*{0.2 cm} \\
\displaystyle \leq -\, \frac32\, \left[ \nu_{ij}\, n_i\, b_{ij} \log \left( \frac{T_{j}}{T_i} \right) + \nu_{ji}\, n_j\, b_{ji} \log \left( \frac{T_{i}}{T_j} \right) \right] = 0\,,
\end{array}
$$
where use has been made of the expression (\ref{aux-temp}) for auxiliary temperatures $T_{ij}$, and of the relation $\nu_{ji}\, n_j\, b_{ji} = \nu_{ij}\, n_i\, b_{ij}$. This completes the proof of the inequality $\mathcal{S}({\bf f}) \leq 0$. Notice that, since each contribution in (\ref{Sf}) has the correct non--positive sign, the space dependence of coefficients $\chi_{ij}({\bf x})$ has not influenced the proof of the entropy dissipation.

Collision equilibria, corresponding to null entropy production $\mathcal{S}({\bf f}) = 0$, are provided by distributions making all non--positive terms in (\ref{Sf}) vanish. Starting from the one--species terms, $\forall i=1, \dots, N$ and $\forall {\bf x} \in \mathbb{R}^3$, if $\chi_{ii}({\bf x}) = 1$ one should have $\langle \mathcal{Q}_{ii}(f_i,f_i), \log f_i \rangle = 0$, otherwise, if $\chi_{ii}({\bf x}) = 0$ one should have $\langle \bar{Q}_{ii}(f_i), \log f_i \rangle = 0$; in both cases this implies $f_i= \mathcal{M}_{ii}$, namely $f_i$ is a Maxwellian distribution depending on species fields $n_i$, ${\bf u}_i$, $T_i$. Passing now to the inter--species interactions, $\forall (i,j)$ and $\forall {\bf x}$ such that $\chi_{ij}({\bf x}) = 1$ one has that $\log f_i$ has to be a collision invariant for the Boltzmann operator $\mathcal{Q}_{ij}(f_i,f_j)$, implying thus ${\bf u}_i = {\bf u}_j$ and $T_i=T_j$ \cite{LucchinKRM,Cercignani2}. In order to have also contributions coming from BGK operators vanishing (in kinetic and also in corresponding macroscopic equations, as discussed in detail in \cite{LucchinKRM}), one should have ${\bf u}_i = {\bf u}_j$ and $T_i=T_j$ also $\forall (i,j)$ and $\forall {\bf x}$ such that $\chi_{ij}({\bf x}) = 0$. In conclusion, admissible collision equilibria are only the expected Maxwellian configurations (\ref{Maxw-eq}) with all species sharing a common velocity ${\bf u}$ and a common temperature $T$.

\section{Hydrodynamic limits}

In this section we derive hydrodynamic Euler--type equations from the hybrid space--dependent Boltzmann--BGK model (\ref{general-mixed}), in three different asymptotic regimes. We consider a binary mixture just for the sake of simplicity, and the relevant kinetic system
\begin{equation}
\begin{array}{l}
\displaystyle \dfrac{\partial f_1}{\partial t}+\mathbf{v}\cdot\nabla_{\mathbf{x}}f_1 = {\sf Q}_{11} + {\sf Q}_{12}\,,\vspace*{0.2 cm}\\
\displaystyle \dfrac{\partial f_2}{\partial t}+\mathbf{v}\cdot\nabla_{\mathbf{x}}f_2 = {\sf Q}_{21} + {\sf Q}_{22}\,,
\end{array}
\label{binary-eq}
\end{equation}
where
$$
{\sf Q}_{ij} = \chi_{ij}({\bf x})\, \mathcal{Q}_{ij}(f_i,f_j) + (1 - \chi_{ij}({\bf x}))\, \bar{Q}_{ij}(f_i)\,,
$$
with $\mathcal{Q}_{ij}(f_i,f_j)$ denoting the Boltzmann operator defined in (\ref{Botlz_term}), and $\bar{Q}_{ij}(f_i)$ the BGK--type relaxation operator (\ref{BGK-model}).
We rescale the equations (\ref{binary-eq}) in terms of a small parameter $\varepsilon$, standing for the Knudsen number (ratio of the particle mean free path to a macroscopic length). We assume different time scales for collisions among particles. Firstly, we consider the classical collision dominated regime, with all collisions playing the dominant role, that leads to classical Euler equations for species densities $n_1$, $n_2$, global mean velocity ${\bf u}$, and global temperature $T$. Then, we investigate a regime where only intra--species collisions are dominant, as it may occur in mixtures of constituents with very disparate masses \cite{Galkin}; in this case we get a multi--velocity and multi--temperature hydrodynamic description. Finally, we consider a situation where the dominant role is played by the heaviest species only, leading to a kinetic--fluid limiting system, where the heavy constituent is described by macroscopic fluid--dynamic equations and the light one by a kinetic equation for its distribution function. In all cases we will expand the distributions in terms of the small parameter $\varepsilon$ as
\begin{equation}
	f_i=f_i^0+\varepsilon f_i^1+\ldots\,,
\label{expansions}
\end{equation}
with the constraints that hydrodynamic variables, corresponding to the collision invariants of the dominant operator, remain unexpanded, as prescribed by the classical Chapman--Enskog asymptotic procedure~\cite{Chapman-Cowling}.

\subsection{Collision dominated regime}

In situations dominated by the whole collision operator, equations (\ref{binary-eq}) rescaled in terms of the Knudsen number $\varepsilon$ read as
\begin{equation}
\begin{array}{l}
\displaystyle \dfrac{\partial f_1}{\partial t}+\mathbf{v}\cdot\nabla_{\mathbf{x}}f_1 = \frac{1}{\varepsilon}\, {\sf Q}_{11} + \frac{1}{\varepsilon}\, {\sf Q}_{12}\,,\vspace*{0.2 cm}\\
\displaystyle \dfrac{\partial f_2}{\partial t}+\mathbf{v}\cdot\nabla_{\mathbf{x}}f_2 = \frac{1}{\varepsilon}\, {\sf Q}_{21} + \frac{1}{\varepsilon}\, {\sf Q}_{22}\,.
\end{array}
\end{equation}
By inserting the expansions (\ref{expansions}), the leading order yields
$$
\left\{
\begin{array}{l}
{\sf Q}_{11}(f_1^0, f_1^0) + {\sf Q}_{12}(f_1^0, f_2^0) = 0\,,\vspace*{0.2 cm}\\
{\sf Q}_{21}(f_2^0, f_1^0) + {\sf Q}_{22} (f_2^0, f_2^0) = 0\,,
\end{array}
\right.
$$
therefore $(f_1^0, f_2^0)$ constitute a collision equilibrium of the whole collision operator that, as proven in the previous section, is always provided by the Maxwellians
\begin{equation}
	f_i^0=n_i\left(\dfrac{m_i}{2\pi T}\right)^{3/2}\exp\left[-\dfrac{m_i}{2T}\left|\mathbf{v}-\mathbf{u}\right|^2\right]\,,\quad i =1,2\,,
	\label{Eulero_ST}
\end{equation}
independently of choosing Boltzmann or BGK operators for the various types of binary encounters, in each position ${\bf x} \in \mathbb{R}^3$.
Collision invariants for the whole Boltzmann--BGK collision operator correspond to species densities, global mean velocity and global temperature, and their evolution equations, obtained using the zero--order approximation (\ref{Eulero_ST}) into the weak form of (\ref{binary-eq}) relevant to collision invariants, are the classical Euler equations
\begin{equation}
	\begin{aligned}
		&\dfrac{\partial n_1}{\partial t}+\nabla_{\mathbf{x}}\cdot(n_1\mathbf{u})=0\,,\\
&\dfrac{\partial n_2}{\partial t}+\nabla_{\mathbf{x}}\cdot(n_2\mathbf{u})=0\,,\\
		&\dfrac{\partial (\rho\mathbf{u})}{\partial t}+\nabla_{\mathbf{x}}\cdot(\rho\mathbf{u}\otimes\mathbf{u}+nT\mathbf{I})={\bf 0}\\
		&\dfrac{\partial }{\partial t}\left(\dfrac12\rho|\mathbf{u}|^2+\dfrac32nT\right)+\nabla_{\mathbf{x}}\cdot\left[\left(\dfrac12\rho|\mathbf{u}|^2+\dfrac52nT\right)\mathbf{u}\right]=0\,,
	\end{aligned}
\end{equation}
with $\mathbf{I}$ denoting the identity matrix. These hydrodynamic conservation equations describe preservation of mass of each component, of global momentum and of total energy in an inert gas mixture.

\subsection{Dominant intra--species collisions}

We consider now a situation where the dominant role in the evolution is played by intra--species collisions only, and the kinetic system (\ref{binary-eq}) is thus rescaled as
\begin{equation}
\begin{array}{l}
\displaystyle \dfrac{\partial f_1}{\partial t}+\mathbf{v}\cdot\nabla_{\mathbf{x}}f_1 = \frac{1}{\varepsilon}\, {\sf Q}_{11} + {\sf Q}_{12}\,,\vspace*{0.2 cm}\\
\displaystyle \dfrac{\partial f_2}{\partial t}+\mathbf{v}\cdot\nabla_{\mathbf{x}}f_2 = {\sf Q}_{21} + \frac{1}{\varepsilon}\, {\sf Q}_{22}\,.
\end{array}
\label{kinet-intra}
\end{equation}
This regime could be meaningful in suitable physical problems relevant to $\varepsilon$--mixtures of heavy and light particles \cite{Galkin}, or in astrophysics \cite{Vranjes}. Leading terms in (\ref{kinet-intra}) provide
$$
{\sf Q}_{11}(f_1^0, f_1^0) = 0\,, \qquad \qquad {\sf Q}_{22}(f_2^0, f_2^0) = 0\,,
$$
hence $f_i^0$ turn out to be Maxwellian distributions depending on species macroscopic fields $n_i$, ${\bf u}_i$, $T_i$:
\begin{equation}
	f_i^0= \mathcal{M}_{ii} =n_i\left(\dfrac{m_i}{2\pi T_i}\right)^{3/2}\exp\left[-\dfrac{m_i}{2T_i}\left|\mathbf{v}-\mathbf{u}_i\right|^2\right]\,,\quad i =1,2\,.
	\label{Eulero_MT}
\end{equation}
Evolution equations of such species fields (corresponding to collision invariants of single--species collision operators) contain, beside the classical Euler--type streaming terms, source terms due to the production of momentum and kinetic energy of inter--species (slower) collisions, described by the operators ${\sf Q}_{12}$ and~${\sf Q}_{21}$. The macroscopic system may be cast as
\begin{equation}
	\begin{aligned}
		&\dfrac{\partial n_1}{\partial t}+\nabla_{\mathbf{x}}\cdot(n_1\mathbf{u}_1)=0\,,\\
        &\dfrac{\partial n_2}{\partial t}+\nabla_{\mathbf{x}}\cdot(n_2\mathbf{u}_2)=0\,,\\
		&\dfrac{\partial (\rho_1\mathbf{u}_1)}{\partial t}+\nabla_{\mathbf{x}}\cdot(\rho_1\mathbf{u}_1\otimes\mathbf{u}_1+n_1T_1\mathbf{I}) = \mathbf{R}_{12}\,,\\
 &\dfrac{\partial (\rho_2\mathbf{u}_2)}{\partial t}+\nabla_{\mathbf{x}}\cdot(\rho_2\mathbf{u}_2\otimes\mathbf{u}_2+n_2T_2\mathbf{I}) = \mathbf{R}_{21}\,,\\
		&\dfrac{\partial }{\partial t}\left(\dfrac12\rho_1|\mathbf{u}_1|^2+\dfrac32n_1T_1\right) +\nabla_{\mathbf{x}}\cdot\left[\left(\dfrac12\rho_1|\mathbf{u}_1|^2+\dfrac52n_1T_1\right)\mathbf{u}_1\right]=  S_{12}\,,\\
&\dfrac{\partial }{\partial t}\left(\dfrac12\rho_2|\mathbf{u}_2|^2+\dfrac32n_2T_2\right) +\nabla_{\mathbf{x}}\cdot\left[\left(\dfrac12\rho_2|\mathbf{u}_2|^2+\dfrac52n_2T_2\right)\mathbf{u}_2\right]=  S_{21}\,,
	\end{aligned}
\label{Euler-MT}
\end{equation}
where $\rho_i = m_i\, n_i$ and exchange terms are provided by
$$
\mathbf{R}_{ij}= \int_{\mathbb{R}^3} m_i\, {\bf v}\, {\sf Q}_{ij} (f_i^0, f_j^0)\, d{\bf v}\,, \qquad \quad
S_{ij}= \int_{\mathbb{R}^3} \frac12\, m_i\, |{\bf v}|^2\, {\sf Q}_{ij} (f_i^0, f_j^0)\, d{\bf v}\,.
$$
Since BGK operators have been built up in \cite{BBGSP} imposing the preservation of Boltzmann binary exchange rates of momentum and energy, contributions ${\bf R}_{ij}$, $S_{ij}$ corresponding to BGK collision operators coincide with the ones obtained with Boltzmann terms, therefore they are independent of the values of coefficients $\chi_{ij}({\bf x}) \in \{0,1\}$ distinguishing Boltzmann and BGK descriptions. More precisely, the (coincident) Boltzmann and BGK exchange rates can be computed explicitly and exactly for Maxwell molecules, for which simple computations provide
\begin{equation}
\begin{array}{ccl}
{\bf R}_{12} & = & \displaystyle \lambda_{12}\,\dfrac{m_1\, m_2}{m_1+m_2}\,n_1n_2(\mathbf{u}_2-\mathbf{u}_1) = -\, {\bf R}_{21}\,, \vspace*{0.2 cm}\\
S_{12} & = & \displaystyle \lambda_{12}\, \dfrac{m_1\,m_2}{(m_1+m_2)^2}\,n_1 n_2 \left[3(T_2-T_1)+(m_1\mathbf{u}_1+m_2\mathbf{u}_2) \cdot (\mathbf{u}_2-\mathbf{u}_1)\right] = -\, S_{21}\,.
\end{array}
\label{R-S}
\end{equation}
For general intermolecular potentials, as explained in Section 2 and in \cite{BBGSP}, a suitable approximation has been introduced in the coefficients $\lambda_{ij}$ in order to allow explicit computations, leading again to the expressions (\ref{R-S}) for the source terms needed in (\ref{Euler-MT}), where now $\lambda_{ij}=\lambda_{ij}(\bar{g})$,  with $\bar{g}$ given in (\ref{bar-g}).

\subsection{Heavy--particles dominated regime}

We consider now a mixture of a heavy constituent (gas 1) and a much lighter one (gas 2), and we assume the mass ratio playing a crucial role in the scaling. A typical physical frame of this kind is provided by plasmas, formed by (heavy) ions and (light) electrons, for which the asymptotic Chapman--Enskog procedure has already been rigorously investigated under different assumptions \cite{GrailleMaginMassot}.
In this subsection we assume that encounters among the heavy particles are assumed dominant ($O(1/\varepsilon)$), interactions involving particles 1 and 2 are supposed occurring at a slower scale ($O(1)$), while collisions among the light ones are almost negligible ($O(\varepsilon)$); thus, the kinetic system turns out to be rescaled as
\begin{equation}
\begin{array}{l}
\displaystyle \dfrac{\partial f_1}{\partial t}+\mathbf{v}\cdot\nabla_{\mathbf{x}}f_1 = \frac{1}{\varepsilon}\, {\sf Q}_{11} + {\sf Q}_{12}\,,\vspace*{0.2 cm}\\
\displaystyle \dfrac{\partial f_2}{\partial t}+\mathbf{v}\cdot\nabla_{\mathbf{x}}f_2 = {\sf Q}_{21} + \varepsilon\, {\sf Q}_{22}\,.
\end{array}
\label{kinet-heavy}
\end{equation}
By inserting expansions (\ref{expansions}), leading order term provides
$$
{\sf Q}_{11}(f_1^0, f_1^0) = 0 \quad \Rightarrow \quad f_1^0=\mathcal{M}_{11} = n_1 \left(\dfrac{m_1}{2\pi T_1}\right)^{3/2} \exp\left[-\dfrac{m_1}{2T_1}\left|\mathbf{v}-\mathbf{u}_1\right|^2\right],
$$
therefore the evolution of the heavy species should be described by means of hydrodynamic equations for $n_1$, ${\bf u}_1$, $T_1$ (collision invariants of ${\sf Q}_{11}$). No constraints for the distribution $f_2$ of the light species arise from the dominant operators in~(\ref{kinet-heavy}). The asymptotic limit of system (\ref{kinet-heavy}) leads thus to fluid--dynamic equations for the macroscopic fields of the heavy gas 1, and to a kinetic equation (the second in (\ref{kinet-heavy}), letting $\varepsilon \to 0$) for the light one. On the right hand side of equations for ${\bf u}_1$ and $T_1$ there appear source terms, already computed in the previous subsection, due to the moments of ${\sf Q}_{12}$ describing inter--species interactions; such production rates of momentum and energy also depend on $n_2$, ${\bf u}_2$, $T_2$, that can be recovered as suitable moments of the unknown distribution $f_2$. In conclusion, our kinetic--fluid system, closed at Euler level accuracy, reads as
\begin{equation}
	\begin{aligned}
		&\dfrac{\partial n_1}{\partial t}+\nabla_{\mathbf{x}}\cdot(n_1\mathbf{u}_1)=0\,,\\
		&\dfrac{\partial (\rho_1\mathbf{u}_1)}{\partial t}+\nabla_{\mathbf{x}}\cdot(\rho_1\mathbf{u}_1\otimes\mathbf{u}_1+n_1T_1\mathbf{I}) = \lambda_{12}\,\dfrac{m_1\, m_2}{m_1+m_2}\,n_1n_2(\mathbf{u}_2-\mathbf{u}_1)\,,\\
		&\dfrac{\partial }{\partial t}\left(\dfrac12\rho_1|\mathbf{u}_1|^2+\dfrac32n_1T_1\right) +\nabla_{\mathbf{x}}\cdot\left[\left(\dfrac12\rho_1|\mathbf{u}_1|^2+\dfrac52n_1T_1\right)\mathbf{u}_1\right] \\
 & = \lambda_{12}\, \dfrac{m_1\,m_2}{(m_1+m_2)^2}\,n_1 n_2 \left[3(T_2-T_1)+(m_1\mathbf{u}_1+m_2\mathbf{u}_2) \cdot (\mathbf{u}_2-\mathbf{u}_1)\right]\,,\\
 & \dfrac{\partial f_2}{\partial t}+\mathbf{v}\cdot\nabla_{\mathbf{x}}f_2 = {\sf Q}_{21}(f_2, \mathcal{M}_{11})\,, \\
	\end{aligned}
\label{Euler-KinFl}
\end{equation}
where the coupling between fluid-dynamic equations and the kinetic one is due both to the appearance of
$$
n_2 = \int_{\mathbb{R}^3} f_2({\bf v})\, d{\bf v}, \quad
{\bf u}_2 = \frac{1}{n_2} \int_{\mathbb{R}^3} {\bf v}\, f_2({\bf v})\, d{\bf v},  \quad
T_2 = \frac{m_2}{3\, n_2} \int_{\mathbb{R}^3} |{\bf v} - {\bf u}_2|^2\, f_2({\bf v})\, d{\bf v}
$$
in the source terms of macroscopic equations, and to the fact that the distribution $\mathcal{M}_{11}(\mathbf{v};n_1,\mathbf{u}_1,T_1)$ of the heavy gas is involved in the Boltzmann or BGK collision operator appearing in the kinetic equation for $f_2$ (light gas).
Some types of kinetic--fluid couplings have been proposed in the literature for various purposes \cite{Cruseillesetal,DegondJinMieussens,GoudonJinetal}, and in this subsection we have shown how a description of this kind may be formally obtained as asymptotic limit of a purely kinetic system.

\section{Conclusions}

We have proved that the hybrid Boltzmann--BGK model for gas mixtures proposed in \cite{LucchinKRM}, where each kind of binary interaction $(i,j)$ may be described by a binary Boltzmann or BGK operator, could also be space--dependent, in the sense that the option of modelling collisions $(i,j)$ by a Boltzmann or by a BGK term could depend on the space position ${\bf x} \in \mathbb{R}^3$ where the particle encounter is occurring. For any choice of coefficients $\chi_{ij}({\bf x}) \in \{ 0,1 \}$ in (\ref{general-mixed}), we have shown that the mixed kinetic model preserves conservations of species densities, global momentum and energy, and fulfills a suitable entropy dissipation allowing to prove that the unique equilibrium configuration is given by Maxwellian distributions depending on densities $n_i$, global mean velocity ${\bf u}$ and global temperature~$T$. This mixed kinetic description could give the possibility to keep an accurate Boltzmann model in regions where rapid phase transitions arise, and to use simpler (but less accurate) BGK operators for collisions occurring in regions where distributions vary less rapidly, close to local equilibrium states. A detailed Boltzmann framework could be preferable to describe the evolution of distribution functions close to the boundaries of the space domain; notice that the treatment of problems defined in half--space or in bounded domains would also require the construction of proper boundary conditions, as for instance in \cite{BardosGolseSone,Lietal}.

We also have investigated the hydrodynamic limit of these hybrid kinetic equations in three different asymptotic scalings, considering for the sake of simplicity a binary mixture. The usual collision dominated regime leads to the well-known Euler equations for global collision invariants, while situations where only intra--species collisions are dominant provide a multi--velocity and multi--temperature description. 
Since, by construction, BGK exchange rates of momentum and energy coincide with the Boltzmann ones, at Euler level the hydrodynamic equations are independent of the type of collision operators (Boltzmann/BGK) used to describe the various collision processes. However, in both regimes we are going on with the Chapman--Enskog asymptotic procedure up to the Navier--Stokes accuracy, and preliminary results provide constitutive relations for pressure tensors and heat fluxes with viscosity and conductivity coefficients actually depending on Boltzmann and/or BGK interaction rates. In order to obtain a more accurate hydrodynamic description, one could assume from the beginning the dominant ($O(1/\varepsilon)$) collision processes modelled with Boltzmann terms, and the remaining $O(1)$ collisions by BGK operators. This would correspond to assume $\chi_{ij} \in \left\{ 1/\varepsilon, 0 \right\}$ and  
${\sf Q}_{ij} = \chi_{ij}({\bf x})\, \mathcal{Q}_{ij}(f_i,f_j)$ $+\, (1 - \varepsilon\, \chi_{ij}({\bf x}))\, \bar{Q}_{ij}(f_i)$.
	
Finally, for the binary mixture we have studied a situation where only the collisions among the heaviest particles play the dominant role, and we have obtained a kinetic--fluid system with macroscopic equations for the heavy species and a kinetic equation for the light one. This regime deserves a deeper investigation in the future, even in presence of more than two species, in order to rigorously prove the derivation of kinetic--fluid equations, commonly used in two--phase flows, as hydrodynamic limit of a purely kinetic description.

\medspace

\paragraph{Acknowledgements.}

The authors are members of INdAM--GNFM, and acknowledge the support from COST Action CA18232 MAT-DYN-NET.
The work has been performed in the frame of the project PRIN 2022 PNRR {\it Mathematical Modelling for a Sustainable Circular Economy in Ecosystems} (project code P2022PSMT7, CUP D53D23018960001) funded by the European Union - NextGenerationEU PNRR-M4C2-I 1.1 and by MUR-Italian Ministry of Universities and Research.
The authors MB and MG also thank the support of the University of Parma through the action Bando di Ateneo 2022 per la ricerca, co-funded by MUR-Italian
Ministry of Universities and Research - D.M. 737/2021 - PNR - PNRR - NextGenerationEU (project {\it Collective and Self-Organised Dynamics: Kinetic and Network Approaches}), and of the PRIN 2020 project {\it Integrated Mathematical Approaches to Socio–Epidemiological Dynamics} (Prin 2020JLWP23, CUP: E15F21005420006).

% ---- Bibliography ----
%

\end{document}